# Successive spatial symmetry breaking under high pressure in the spin-orbit-coupled metal $Cd_2Re_2O_7$

Jun-ichi Yamaura,[1] Keiki Takeda,[2] Yoichi Ikeda,[3] Naohisa Hirao,[4] Yasuo Ohishi,[4] Tatsuo C. Kobayashi,[3] Zenji. Hiroi[5]

[1]Materials Research Center for Element Strategy, Tokyo Institute of Technology, Yokohama, Kanagawa 226-8503, Japan

[2]Muroran Institute of Technology, Muroran, Hokkaido 050-8585, Japan

[3]Graduate School of Natural Science and Technology, Okayama University, Okayama 700-8530, Japan

[4]Japan Synchrotron Radiation Research Institute (JASRI)/SPring-8, Sayo-gun, Hyogo 679-5198

[5]Institute for Solid State Physics, University of Tokyo, Kashiwanoha, Kashiwa, Chiba 277-8581

The $5d$-transition metal pyrochlore oxide $Cd_2Re_2O_7$, which was recently suggested to be a prototype of the spin-orbit-coupled metal [*Phys. Rev. Lett.* **115**, 026401 (2015)], exhibits an inversion-symmetry breaking (ISB) transition at 200 K and a subsequent superconductivity below 1 K at ambient pressure. We study the crystal structure at high pressures up to 5 GPa by means of synchrotron X-ray powder diffraction. A rich structural phase diagram is obtained, which contains at least seven phases and is almost consistent with the electronic phase diagram determined by previous resistivity measurements. Interestingly, the ISB transition vanishes at ~4 GPa, where the enhancement of the upper critical field was observed in resistivity. Moreover, it is shown that the point groups at 8 K, probably kept in the superconducting phases, sequentially transform into piezoelectric, ferroelectric, and centrosymmetric structures on the application of pressure.

In recent years, there has been a growing interest in the significant role of spin-orbit coupling (SOC) in condensed matter physics. Pyrochlore oxides with heavy 5$d$-transition metals in the pyrochlore lattice made of corner-sharing tetrahedra are candidate materials for the platform of studying new aspects of electronic, magnetic and structural properties strongly related to SOC [1,2]. In $R_2Ir_2O_7$ (R = rare earth element) and $Cd_2Os_2O_7$, probably due to strong SOC and electron correlations, the time-reversal symmetry is spontaneously broken to cause all-in/all-out magnetic orders and metal-insulator transitions simultaneously, while the spatial inversion symmetry is preserved during the transitions [3−12]. In sharp contrast, $Cd_2Re_2O_7$ exhibits an inversion-symmetry breaking (ISB) transition with time-reversal symmetry and a subsequent superconducting transition [13−23].

Liang Fu recently suggested that $Cd_2Re_2O_7$ is a potential material of the spin-orbit-coupled metal (SOCM) with a large SOC and inversion symmetry [24]. He pointed out a novel type of Fermi liquid instability in the SOCM, resulting in a variety of time-reversal-invariant and parity-breaking phases whose Fermi surface is spontaneously deformed and spin split: they are gyrotropic, ferroelectric, and multipolar orders in terms of symmetry. Moreover, Kozii and Fu, and Wang *et al.* proposed the possibility of odd-parity superconductivity meditated by parity fluctuations in the vicinity of ISB in the SOCM [25,26].

$Cd_2Re_2O_7$ has a semimetallic band structure mainly derived from Re 5$d$-electrons at room temperature [27−29]. It undergoes a successive structural transition from high-temperature cubic *Fd*−3*m* (phase I) to tetragonal *I*−4*m*2 (phase II) at $T_{s1}$ = 200 K and to tetragonal *I*4$_1$22 (phase III) at $T_{s2}$ = 120 K [13-19]. The $T_{s1}$ transition of the second-order type is considered as a band Jahn−Teller transition lifting the highly degenerate bands, because there are large changes in the electronic properties in spite of the tiny tetragonal distortion at $T_{s1}$ [23,30] On the other hand, the origin of the $T_{s2}$ transition, which is of the first-order, is thus far unclear [15,31].

Effects of applying high pressure on $Cd_2Re_2O_7$ have been examined. In early study, a suppression of the $T_{s1}$ transition was observed at ~3.5 GPa [32,33]. More precise measurements of electrical resistivity by Kobayashi *et al.* found as many as eight electronic phases in the pressure range up to 5 GPa [34]. Furthermore, Malavi *et al.* reported a metal-nonmetal transition accompanied by a cubic–rhombohedral structural transition at 14 GPa at room temperature [35]. The reason why so many phases appear under high pressure remains unknown, but must be related to the Fermi liquid instabilities of the SOCM. To get an insight into the electronic instabilities of $Cd_2Re_2O_7$, it is crucial to investigate its crystal structures at high pressures and low temperatures.

Cd$_2$Re$_2$O$_7$ exhibits superconductivity at $T_c$ = 1.0 K in phase III in the absence of the spatial inversion symmetry [20−22]. Such non-centrosymmetric (NCS) superconductors have attracted much attention, because they can yield exotic pairing states with spin-singlet and triplet mixtures [36]. For example, CePt$_3$Si and CeRhSi$_3$ have been well characterized, which show a large upper critical magnetic field $B_{c2}$ exceeding the Pauli limiting field as an evidence of exotic superconductivity [36,37]. We stress that Cd$_2$Re$_2$O$_7$ loses inversion symmetry as a result of phase transition with temperature in contrast to the fact that the hitherto known NCS superconductors lack their inversion symmetry throughout the temperature range; one would expect fluctuation-related phenomena only for the former. However, the superconductivity of Cd$_2$Re$_2$O$_7$ at ambient pressure has been established as a conventional $s$-wave type [20−23,30,38,39]. On the other hand, applying pressure to this material enhances $T_c$ up to 3 K at 2 GPa [34]. Insights into potentially exotic superconductivity have been obtained from the results of the enhanced $B_{c2}$ over the Pauli limiting and the electronic-mass enhancement near a critical pressure of ~4 GPa [34].

In this paper, the crystal structures of Cd$_2$Re$_2$O$_7$ are investigated up to ~5 GPa by means of synchrotron X-ray powder diffraction. It is revealed that most of the multiple high-pressure phases previously observed in resistivity match up with crystallographic phases of distinct symmetry and tiny structural modifications. Moreover, it is found that the ISB line disappears at ~4 GPa, where $B_{c2}$ is enhanced.

Powder samples of Cd$_2$Re$_2$O$_7$ were prepared from CdO and ReO$_3$ with a molar ratio of 1:1 in a sealed quartz tube at 700 °C for 5 days. The particle size of the powder sample was equalized by sedimentation in toluene. We employed synchrotron radiation of wavelength λ = 0.4217 Å for high-pressure X-ray diffraction at BL-10XU, SPring-8. Two-dimensional images were collected on an imaging plate in RIGAKU R-AXIS, and integrated to yield 2θ-intensity data by the WIN-PIP program. Pressure up to ~5 GPa was generated using a diamond anvil cell (DAC) composed of a pair of diamond anvils with 800 μm culet and a SUS301T gasket with a 400 μm hole in between. A mixture of MeOH-EtOH in the ratio 4:1 was used as a pressure transmitting medium, and the pressure was determined by an in-situ measurement of ruby fluorescence [40]. The pressure was set to 1, 3, 4, 5 GPa at ~250 K by applying He gas pressure to the DAC equipped with a closed-cycle refrigerator, and the DAC was cooled down to the lowest temperature ~8 K. Then the diffraction data were collected upon heating the DAC to ~250 K; the pressure was slightly decreased by ~0.2 GPa at elevated temperatures.

We firstly focus on the presence of inversion symmetry in the crystal. A good indicator for the loss of inversion center is the forbidden reflection of 00$l$ ($l$ = 4$n$+2) in the original space group of $Fd\overline{3}m$, which appears by the loss of the $d$-glide plane and the $4_1$-screw axis symmetries [14,15]. Figures 1(a) and 1(b) show the X-ray profiles of the 002 reflection at 3 and 5 GPa, respectively; all the indices are based on the cubic lattice system. The 002 reflection at 3 GPa appears in the temperature range of 102–141 K, whereas at 5 GPa, it is not detected in any part of the temperature range within the experimental error.

Figure 2 shows the temperature dependences of the integrated intensity of the 002 reflection $I(002)$. $I(002)$ develops gradually below ~180 K at 1 GPa [Fig. 2(a)], which coincides with $T_{s1}$ = 180 K obtained from the resistivity measurement [34]. This means that the cubic–tetragonal transition ($T_{s1}$ = 200 K at ambient pressure) progressively shifts to lower temperatures on the application of pressure [34]. On further application of pressure, $I(002)$ appears below $T_{s1}$ = 140(1) K at 3 GPa and 125(1) K at 4 GPa, and is absent at 5 GPa, as plotted in Figs. 2(b)−2(d). Accordingly, the ISB is suppressed by pressure, and eventually vanishes at a critical pressure between 4 and 5 GPa [41]. The continuous change of $I(002)$ indicates that the $T_{s1}$ transition remains second-order below 4 GPa. Furthermore, we observed discontinuous changes in $I(002)$ at $T_{s3}$ ~ 55 K (3 GPa) and at $T_{s3}'$ ~ 55 K (4 GPa).

We next give an account of lattice distortions, which can be seen by investigating the fundamental $h00$ and $hhh$ reflections. In general, it is often the case that $h00$ ($hhh$) peaks split, whereas $hhh$ ($h00$) peaks do not when a crystal deforms from cubic to tetragonal or orthorhombic (hexagonal or trigonal). Moreover, the splitting of both the peaks indicates monoclinic or triclinic distortions. Figures 1(c)−1(f) show the X-ray profiles of the 0016 and 888 reflections at 3 and 5 GPa. At 3 GPa and 102 K below $T_{s1}$, neither 0016 nor 888 reflection indicates discernible broadening. In the present experimental setup, a tetragonality of $\sqrt{2}a_t/c_t - 1$ as large as $2 \times 10^{-4}$ can be detected, where the suffix t denotes the lattice constant in the tetragonal setting. This value is smaller than that of phase II at ambient pressure, $4 \times 10^{-4}$ at 120 K [14]. Thus, it is likely that the crystal converts to a NCS cubic or tetragonal phase with an undetectable tetragonality. Taking into account that similar behavior in resistivity has been observed at $T_{s1}$ up to ~4 GPa, we suggest that tetragonal phase II with smaller tetragonalities is preserved under pressure below $T_{s1}$.

Small but distinct peak broadenings are observed in the 0016 and 888 reflections simultaneously at 81 K and 3 GPa. This indicates that the crystal system is either monoclinic or triclinic; the tetragonal phase III appearing below $T_{s2}$ = 120 K at ambient pressure seems to already disappear. On the other hand, at 5 GPa, both the 0016 and 888 peaks broaden at 80 K, implying that the crystal system is either monoclinic or triclinic in the presence of inversion symmetry.

Figure 2 shows the temperature dependences of the broadening parameters $\Delta H$ for the 0016 and 888 reflections. The $\Delta H$ is estimated by subtracting the full width at half maximum (FWHM) at room temperature and 1 GPa from the width at each point measured; the FWHM is obtained by fitting a profile to a single pseudo-Voigt function. $\Delta H(0016)$ and $\Delta H(888)$ start broadening simultaneously below $T_{s5}$, indicating second-order transitions [Figs. 2(a)–2(c)]: $T_{s5} = $ ~130, 104.8(8), and 125(1) K at 1, 3, and 4 GPa, respectively. As mentioned above, the tetragonal phase II may exist between $T_{s1}$ and $T_{s5}$ at 1 and 3 GPa, while is almost absent at 4 GPa; $T_{s1}$ and $T_{s5}$ appear coincidentally there.

At lower temperatures, $\Delta H(0016)$, but not $\Delta H(888)$, exhibits discontinuous changes at ~55 K under 3 and 4 GPa, which may correspond to the anomalies of $I(002)$ at $T_{s3}$ and $T_{s3}'$, respectively. There are distinct anomalies in resistivity at nearly the same temperatures. Thus, there must be additional first-order transitions at $T_{s3}$ and $T_{s3}'$. At further low temperatures, $\Delta H(0016)$ and $\Delta H(888)$ rapidly increase, suggesting that the lattice distortions have either changed or accelerated. Another structural transition or a development of electron–lattice coupling may exist at low temperatures. However, as there is no corresponding anomaly in resistivity, more sophisticated study is required to elucidate the origin of these changes.

All the X-ray diffraction profiles below $T_{s5}$ and $T_{s3}$ can be well reproduced by monoclinic lattices as typically shown in the X-ray profile at 7.7 K and 3.1 GPa in Fig. S1 [42]. Thus, it is plausible that all the low-temperature phases below $T_{s5}$ are monoclinic, which means that the transitions at $T_{s5}$ and $T_{s3}$ ($T_{s3}'$) are tetragonal–monoclinic and monoclinic–monoclinic, respectively. The lattice distortions are estimated by the quantities $a_m/b_m - 1 = 6.1 \times 10^{-4}$ and $1.6 \times 10^{-3}$, and $\sqrt{2}/2(a_m+b_m)/c_m - 1 = -2.0 \times 10^{-3}$ and $-3.2 \times 10^{-3}$ at 61.1 K (3.3 GPa) and at 7.7 K (3.1 GPa), respectively, where the suffix m denotes the lattice constant in monoclinic setting [43]. These are extremely small in comparison with the structural distortion of the other pyrochlore oxide $Hg_2Ru_2O_7$ in its antiferromagnetic insulating phase: $\sqrt{2}/2(a_m+b_m)/c_m - 1 = -1.4 \times 10^{-2}$ [44].

At 5 GPa, where the crystal remains centrosymmetric at all temperatures, both $\Delta H(0016)$ and $\Delta H(888)$ start broadening below $T_{s4}$ ~ 110 K, and then exhibit a discontinuous change at $T_{s3}''$ ~ 50 K [Fig. 2(d)]. The profiles below $T_{s4}$ and $T_{s3}''$ can be fitted by monoclinic systems with tiny lattice distortions: $a_m/b_m - 1 = 1.4 \times 10^{-4}$ and $1.8 \times 10^{-3}$, and $\sqrt{2}/2(a_m+b_m)/c_m - 1 = -2.6 \times 10^{-3}$ and $-2.7 \times 10^{-3}$ at 60.6 K (5.2 GPa) and 8.0 K (5.2 GPa), respectively [43]. Accordingly, $T_{s4}$ and $T_{s3}''$ are regarded as cubic–monoclinic and monoclinic–monoclinic transitions, respectively. As can be inferred from the present results along with the previous resistivity measurements, which show a continuous transition at $T_{s4}$ and a transition with a thermal hysteresis at $T_{s3}$", we conclude that the $T_{s4}$ transition is of the second-order and the $T_{s3}$" transition is of the first-order. The finite value of $\Delta H$ above $T_{s4}$ is seemingly due to extrinsic effects such as inhomogeneity in the distribution of pressure inside the DAC.

The present results are summarized in the temperature–pressure phase diagram [Fig. 3(a)], along with the results from the previous resistivity measurements [34]. Filled and open marks are obtained from the structural and resistivity measurements, respectively, and second- and first-order transition lines are shown by thin and thick lines, respectively. The names of the phases in Roman numerals are from the previous study. First, the ISB transition temperatures based on the 002 forbidden reflection coincide with the $T_{s1}$ line from resistivity. Although the following transition line $T_{s1}'$ below ~100 K was not passed through in the present experiment, we naturally expect the $T_{s1}$–$T_{s1}'$ line as the ISB boundary. Note that the ISB transition changes its character from the second-order type at $T_{s1}$ to the first-order type at $T_{s1}'$, and then vanishes at around 4.3 GPa.

The apparent difference from the previous phase diagram is the presence of a new monoclinic phase below the second-order transition at $T_{s5}$, which was not detected in resistivity and is defined as phase IX in the present study. This phase survives down to low temperature, while above 2.8 GPa, it transforms into phase IV or VII at $T_{s3}$ or $T_{s3}'$ ~ 55 K, as observed both in structure and resistivity. At higher pressures above 4.3 GPa, there exist two structural and electronic transitions at $T_{s4}$ and $T_{s3}$"; phases V and VIII take different structures.

The previous resistivity study has determined three vertical lines at the lowest temperature in the phase diagram, which separate phases IX, IV, VII and VIII, from discontinuous changes in $T_c$, coefficient $A$ of $T^2$, and the residual resistivity [34]; all the transitions are of the first-order. The present study can detect all of them except the boundary between phases IV and VII, which is indistinguishable from the structural perspective. This is also the case for the line separating phases IX and VI above $T_{s3}$ / $T_{s3}'$. This fact suggests that the differences between these phase pairs may stem purely from electronic reasons with less coupling to the lattice. Otherwise, phases VI and VII may correspond to mixed-phase regions between the adjacent phases.

We now move to assess the space groups of the high-pressure phases. As the supercell reflection is unobserved, the candidates of the space groups are considered as non-isomorphic $t$ (translationengleiche) subgroups from the $Fd$–$3m$ prototype structure of pyrochlore oxides [Fig. 3(b)][45]. The Landau theory validates the holding of group-subgroup relations for a second-order transition, while, for a first-order transition, it is often the case that the group-subgroup relation is unsatisfied. Indeed, the second-order I–II transition and first-order II–III transition in $Cd_2Re_2O_7$ are consistent with these criteria; in the group-subgroup map of Fig. 3(b), $I\bar{4}m2$ for phase II is in the downstream of $Fd\bar{3}m$ for phase I, while $I4_122$ for phase III is in another branch. We proceed to identify the space groups according to these criteria. As all the high-pressure phases IV–IX are identified as monoclinic systems, the potential space groups without inversion symmetry are $C2$, $Cm$ and $Cc$, belonging to polar point groups, whereas those with inversion symmetry are $C2/m$ and $C2/c$.

The tetragonal phase below $T_{s1}$ at 1–4 GPa is regarded as phase II ($I\bar{4}m2$), and phase III ($I4_122$) may disappear at low pressure. Since the transition from phase II to phase IX at $T_{s5}$ is of the second-order, phase IX is ascribed to either $Cm$ or $C2$, a subgroup of $I\bar{4}m2$ [Fig. 3(b)]. On the one hand, phase V, neighboring to phase IX (VI) separated by the first-order line, is either of $C2/m$ and $C2/c$. Thus, $C2$, which is a subgroup of $C2/m$ and $C2/c$, is excluded for phase IX (VI). Thus, the space group of phase IX (VI) is uniquely determined as $Cm$. This fact also means that phase V is $C2/c$, and automatically the lower-temperature phase VIII is $C2/m$. This is consistent with the second-order transition from phase I to V and the first-order transition to phase VIII. Moreover, because of the first-order separation between the phases IV (VII) and VIII as evidenced by resistivity, the phase IV (VII) becomes $Cc$, which is separated from phase IX ($Cm$) by the first-order line, as expected from the group-subgroup relation. Therefore, all the monoclinic space groups are uniquely determined so as to satisfy the Landau relation and the related assumption. Thus, the determined space groups are described in the phase diagram of Fig. 3(a). For further verification, the observations of the soft phonon modes from Raman scattering experiments under pressure would be helpful.

The multiphases of $Cd_2Re_2O_7$ are thus linked by tiny structural changes as well as electronic properties including superconductivity. This strongly suggests that the phase transitions originate from electronic instability in the SOCM [24]. It is intriguing to note that the point groups (space groups) at the lowest temperature of 8 K transform successively as 422 ($I4_122$), $m$ ($Cm$), $m$ ($Cc$), and $2/m$ ($C2/m$) with pressure. The superconductivity occurs below $T_c \sim 3$ K probably in the same structures, because no anomalies in resistivity were observed down to $T_c$. In other words, the present material sequentially crystallizes into piezoelectric, ferroelectric and centrosymmetric structures. The relation between these and the symmetry-broken phases for the SOCMs should be examined in future studies.

Concerning superconductivity, it is likely that such an exotic pairing state as predicted in Refs. 25 and 26 are realized, because the significant enhancement of $B_{c2}$ over the Pauli limiting [34] occurs near the ISB boundary. In order to get insight on the electronic instability of the SOCM and the mechanism of superconductivity, it would be intuitive to look at the possible displacements of the heavy element Re [42,45]. A more detailed discussion awaits the determination of the actual atomic displacements, which will be carried out in the near future.

In summary, the structures of the 5*d*-pyrochlore oxide superconductor $Cd_2Re_2O_7$ were investigated by means of synchrotron X-ray powder diffraction under high pressures up to ~5 GPa. We showed that there are a number of structural transitions with extremely small distortions, which are mostly consistent with the anomalies reported in the previous resistivity measurements. The present results support that the compound can be viewed as a SOCM having Fermi liquid instabilities to various parity-breaking phases as well as exotic superconductivity near ISB.

We are grateful to Y. Kuramoto, T. Hasegawa, and H. Harima for their fruitful discussion. The synchrotron radiation experiments were performed with the approval of JASRI (No. 2011A1184). This work was supported by MEXT Elements Strategy Initiative to Form Core Research Center and also by JSPS KAKENHI (No. 16K05434).

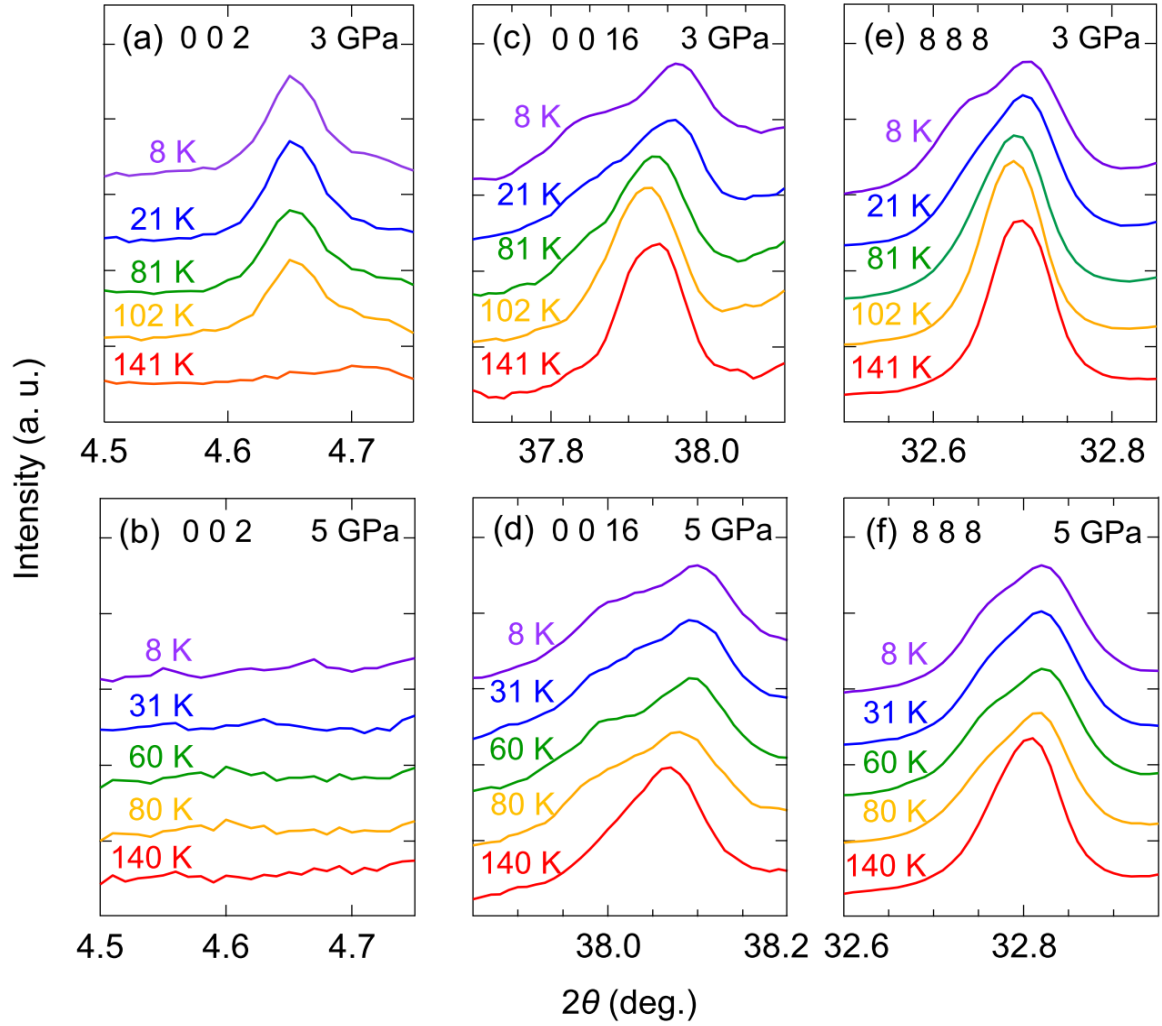

FIG. 1. Temperature evolutions of representative X-ray diffraction profiles under pressures of 3 and 5 GPa: (a) and (b) for the forbidden 002 reflections, (c) and (d) for the 0016 reflection, and (e) and (f) for the 888 reflections, respectively. The indices are based on the cubic lattice structure of phase I.

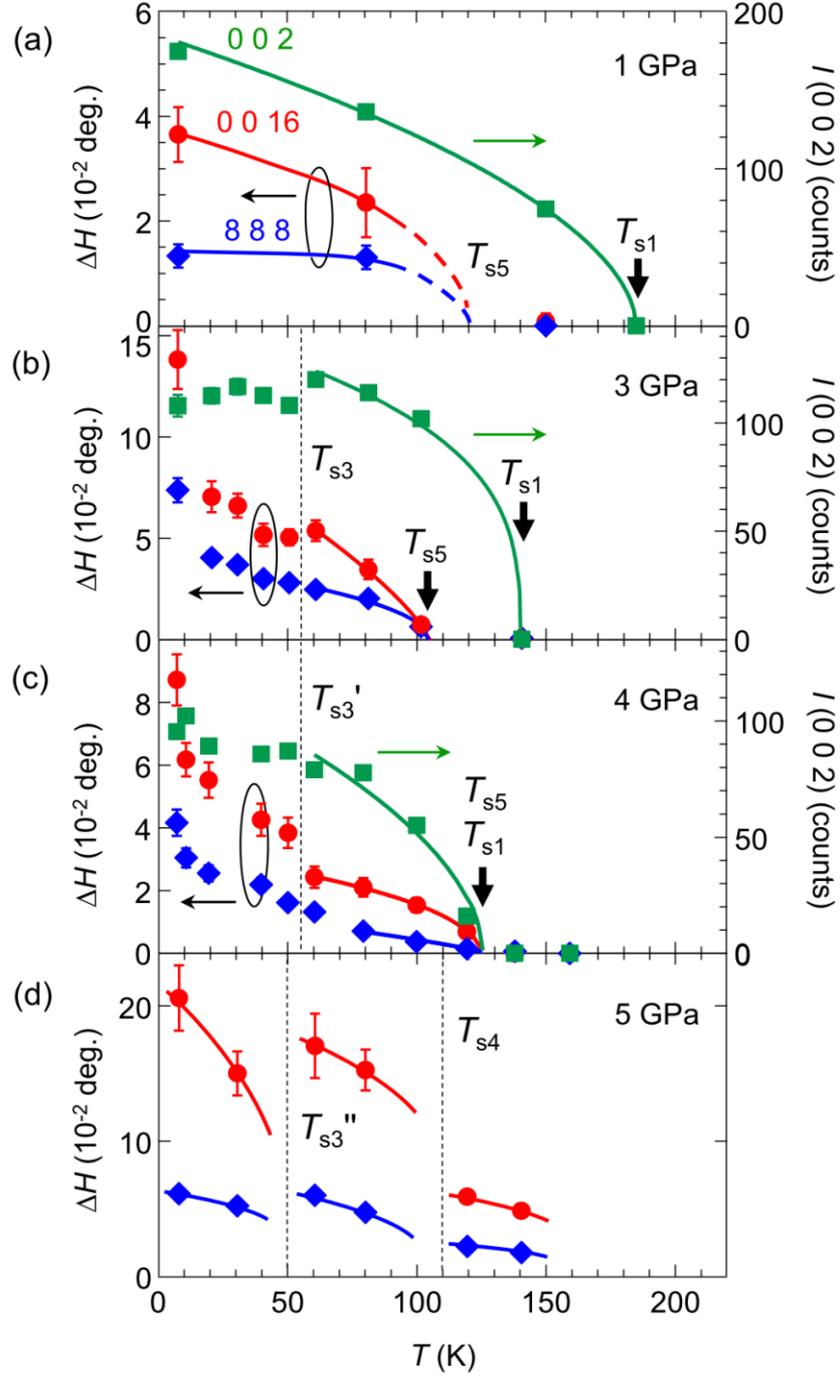

FIG. 2. Temperature dependence of the intensities of the forbidden 002 reflection $I(002)$, and the broadening parameters $\Delta H$ for the 0016 (red) and 888 (blue) reflections at 1 GPa (a), 3 GPa (b), 4 GPa (c), and 5 GPa (d). $I(002)$ is normalized by the intensity of the fundamental 111 reflection. $\Delta H$ is estimated by subtracting the FWHM at room temperature and 1 GPa from that at each point. The FWHM is obtained by fitting to a single pseudo-Voigt function. The lines are fits to the equation $(T_c - T)^\beta$ except for those at 5 GPa, which are drawn for eye-guides.

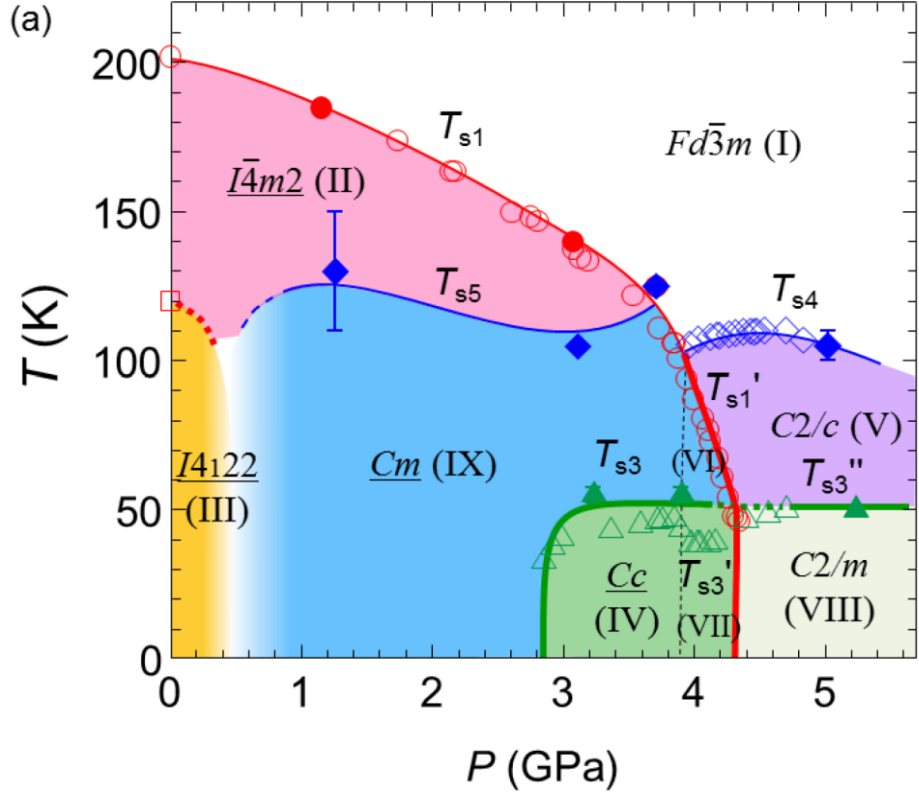

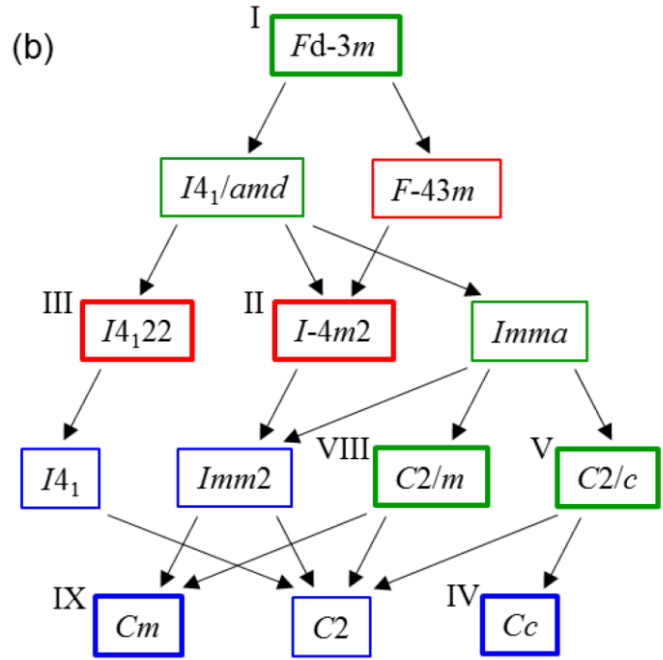

FIG. 3. (a) Temperature-pressure phase diagram of $Cd_2Re_2O_7$. Filled marks indicate the transition temperatures obtained in the present study: transitions with ISB (red), monoclinic distortions (blue), and discontinuous changes in $\Delta H(0016)$ (green). Open marks refer to the anomalies in the previous resistivity measurements [34]. The pressure values of the resistivity

data have been slightly corrected, taking into account the reduction in pressure at elevated temperatures (no in-situ measurements of pressure was done there). Thin and thick lines represent second-order and first-order transitions, respectively. Space groups are shown for each phase; those without inversion symmetry are underlined. Phases IV and VII, or phases IX and VI are indistinguishable from our structural measurements but are distinguishable from the resistivity data. (b) Relevant translationengleiche subgroups from the $Fd-3m$ prototype structure of the pyrochlore oxide. Space groups in the green, red, and blue frames have centrosymmetric, piezoelectric, and ferroelectric point groups, respectively. Those assigned to the multiple phases of $Cd_2Re_2O_7$ are marked by thick frames.